\documentclass[12pt]{iopart}
\usepackage{graphicx}
\usepackage{epsfig}
\usepackage{ulem}
\usepackage{morefloats}
\usepackage{rotating,ulem}
\usepackage{color}

\def\beq{\begin{equation}}
\def\eeq{\end{equation}}
\def\beqa{\begin{eqnarray}}
\def\eeqa{\end{eqnarray}}
\def\ban{\begin{eqnarray*}}
\def\ean{\end{eqnarray*}}
\def\bi{\begin{itemize}}
\def\ei{\end{itemize}}

\begin{document}

\title[Experimental determination of chemical constants]
{Improved method for the experimental determination of in-medium
effects from heavy-ion collisions}

\author{Helena Pais$^1$, R\'emi Bougault$^2$, Francesca Gulminelli$^2$, Constan\c ca Provid{\^e}ncia$^1$,Eric Bonnet$^3$, Bernard Borderie$^4$, Abdelouahad  Chbihi$^5$, John D. Frankland$^5$, Emmanuelle Galichet$^{4,6}$, Di\'ego Gruyer$^2$, Maxime Henri$^5$, Nicolas Le Neindre$^2$, Olivier Lopez$^2$, Loredana Manduci$^{2,7}$, Marian Parl\^og$^2$, Giuseppe Verde$^8$ }
\address{$^1$CFisUC, Department of Physics, University of Coimbra,
  3004-516 Coimbra, Portugal.}
\address{$^2$Normandie Univ, ENSICAEN, UNICAEN, CNRS/IN2P3, LPC Caen, F-14000 Caen, France.}
\address{$^3$SUBATECH UMR 6457, IMT Atlantique, Universit\'e de Nantes, CNRS-IN2P3, 44300 Nantes, France.}
\address{$^4$Universit\'e Paris-Saclay, CNRS/IN2P3, IJCLab, 91405 Orsay, France.}
\address{$^5$Grand Acc\'el\'erateur National d'Ions Lourds (GANIL), CEA/DRF-CNRS/IN2P3, Bvd. Henri Becquerel, 14076 Caen, France.}
\address{$^6$Conservatoire National des Arts et Metiers, F-75141 Paris Cedex 03, France.}
\address{$^7$Ecole des Applications Militaires de l'Energie Atomique, BP 19 50115, Cherbourg Arm\'ees, France.}
\address{$^8$INFN - Sezione Catania, via Santa Sofia 64, 95123 Catania, Italy.}
\eads{\mailto{hpais@uc.pt}, \mailto{bougault@lpccaen.in2p3.fr}, \mailto{gulminelli@lpccaen.in2p3.fr}, \mailto{cp@uc.pt}}

\begin{abstract}

The equation of state with light clusters for nuclear and stellar matter is determined using chemical equilibrium constants evaluated from the analysis of the recently published (Xe$+$Sn) heavy ion data, corresponding to three reactions with different isotopic contents of the emission source. The measured multiplicities are used to extract the thermodynamic properties, and an in-medium correction to the ideal gas internal partition function of the clusters is included in the analysis. This in-medium correction and its respective uncertainty are calculated via a Bayesian analysis, with the unique hypothesis that the different nuclear species in a given sample must correspond to a unique common value for the density of the expanding source. 
Different parameter sets for the correction are tested, and the effect of the radius of the clusters on the thermodynamics and on the chemical equilibrium constants is also  addressed. It is shown that the equilibrium constants obtained are almost independent of the isospin content of the analysed systems.
Finally, a comparison with a relativistic mean field model proves that data are consistent  with a universal in-medium correction of the scalar $\sigma$-meson coupling for nucleons bound in clusters. The obtained value,  $g_s/g_s^0 = 0.92 \pm 0.02$, is larger than that obtained in a previous study not including in-medium effects in the data analysis. This result implies a smaller effect on the binding energy of the clusters and, as a consequence, larger melting densities, and an increased cluster contribution in supernova matter.

\end{abstract}

\maketitle

\section{Introduction} \label{sec:intro}

Light and heavy clusters are supposed to form in three main astrophysical sites: neutron star
mergers, core-collapse supernovae and the inner crust of neutron stars
\cite{Sumiyoshi2008,Fischer2014,Furusawa2013,Furusawa2017,Arcones2008}.
 These clusters affect the neutrino mean free path and, as a consequence, the transport properties of the system. Shortly after the supernova explosion, the presence of light clusters is expected to reduce the fraction of free protons \cite{Sumiyoshi2008,Arcones2008} and affect the  weak processes involving the diffusion of neutrinos \cite{Arcones2008}. This is confirmed by recent calculations \cite{Fischer2016,Fischer2017}, even if   the effect on the protoneutron star evolution seems weaker  than the results in \cite{Arcones2008} indicated. Light clusters also play a role in the NS merger evolution, in particular, in the viscous evolution of the disk and the initial conditions for r-process nucleosynthesis \cite{Rosswog2015}. A recent discussion on how light clusters  can influence  the dynamics of astrophysical processes can be found in \cite{Oertel2017}.

Properties of light clusters are modified in a dense medium. There are  some works  that study the effect of the medium on these clusters from a quantum-statistical approach \cite{Roepke2015}. Phenomenological models, on the other hand, have the in-medium interactions introduced via coupling constants, which should be fixed with the help of experimental constraints \cite{HempelPRC91,PaisPRC97}.

In heavy-ion collisions these clusters may also form. 
 Even though the conditions may be different, heavy ion experiments can provide 
useful constraints to settle the in-medium modifications of cluster properties. Such modifications can then be implemented in theoretical models adapted to the astrophysical context.
In particular, a very interesting data selection and analysis technique was proposed by the  NIMROD collaboration \cite{QinPRL108} to extract chemical constants from heavy ion data, an observable which can benchmark the theoretical calculations of clusters in the medium. 
So far we had only one constraint from experimental data \cite{QinPRL108}. Recently, the same analysis was applied by the INDRA collaboration on different sets of data \cite{BougaultJPG19}. 
These data confirm the low density values as obtained in Ref.~\cite{QinPRL108}, though particle production is not directly comparable with the one of Ref.~\cite{QinPRL108}, because the measured temperatures are not the same. However, it was pointed out in \cite{BougaultJPG19} that in the analysis of  Ref.~\cite{QinPRL108} 
 an ideal gas partition sum is supposed in order to extract the system density, which is not consistent with the fact that the  experimental chemical constants are not reproduced by the ideal gas hypothesis.
To resolve this contradiction, we revisit the analysis of Ref.~\cite{BougaultJPG19}, by explicitly allowing for the possibility of in-medium effects. 

The main results of this new analysis were already reported in a previous work \cite{Pais20-PRL}. 
 In the present paper, we intend to  explain in detail the different steps of this new analysis. 

Our new analysis goes beyond the ideal gas hypothesis, because we
include in-medium effects via a correction to the internal partition
function. 
 Specifically, the ideal gas expression relating the measured abundances to the average volume is  
modified, and the a-priori unknown modification is treated as a random variable with a pdf determined 
via Bayesian tools. A flat non-informative prior is assumed, and the posterior is obtained imposing the principle that, for a given set of chemical constants to represent a well defined thermodynamic condition $(T,\rho,y_p)$
and be comparable to an equilibrium thermal model, the average volume derived by the momentum space density power law \cite{DasGuptaPR72} should be the same for all the different cluster species present in the thermodynamic ensemble. 

With this correction, the thermodynamic conditions of the experimental data, i.e. the temperature $T$, density $\rho$ and global proton fraction $y_p$, were constructed, and finally, we were able to  calculate the experimental chemical equilibrium constants consistently, and estimate quantitatively the systematic uncertainty due to the volume estimation.

From the theoretical point of view, in Ref.~\cite{PaisPRC97}, the authors proposed a model to include
in-medium effects in the equation of state for low density
non-homogeneous warm stellar matter. The model includes a) an effective mass shift term to
the total binding energy of the clusters derived in the
Thomas-Fermi approximation, which mimics the exclusion volume approach;
b) a modification to the scalar cluster-meson couplings, fitted
to the Virial EoS \cite{virial}. In that work, four light clusters were considered, $^4$He,
$^3$He, $^3$H, and $^2$H, and the predictions of the model were compared with
the experimental constraints that existed at the time, the chemical
equilibrium constants from Ref.~\cite{QinPRL108}. 
Universal coupling constants for the scalar
cluster-meson field were extracted from this comparison, which can  be very easily implemented in other Relativistic Mean Field EoS, that may ultimately turn out to be more
adequate to reproduce nuclear matter properties. Later, the authors
included heavier light clusters, together with a heavy cluster (pasta
configuration) \cite{PaisPRC2019}. The motivation for these
calculations was the fact that indeed both in astrophysical sites as
well in HIC experiments, heavy clusters should  also  form, and, therefore,
should be included in the EoS. The new model predictions also
reproduced quite well the experimental data from Ref.~\cite{QinPRL108} with the same fitted value of the cluster coupling.

Those results are revisited in the present work. We perform a comparison of  the theoretical
model of  Ref.~\cite{PaisPRC97} with the new analysis of the experimental data   obtained by the INDRA
collaboration \cite{BougaultJPG19}. We will show that we need a larger scalar cluster-meson coupling, as compared to the one that was used to fit the data of Ref.~\cite{QinPRL108}. 

In a future work, it would be extremely interesting to revisit the analysis of the experimental data of Ref.~\cite{QinPRL108}, in the same spirit as the one presented in this paper, such as to check if the two data sets are compatible. 
Indeed this would be a very strong evidence that the short-lived, expanding nuclear sources formed in heavy-ion collisions can be treated as a collection  of statistical ensembles, which for the moment is only a working hypothesis of all the different analyses.

The plan of the paper is as follows.
In Section \ref{sec:form}, we present the formalism used, in Section \ref{sec:analysis} the determination from a Bayesian calculation of the correction to the internal partition function is explained. The chemical equilibrium constants are calculated  in Section \ref{sec:kci} and the different hypotheses are critically analyzed in Sections \ref{sec:correction} and \ref{sec:radius}. The data are compared to the theoretical model in section \ref{sec:model}. Finally, in Section \ref{sec:conc} some conclusions are drawn.

\section{Formalism} \label{sec:form}

Equilibrium chemical constants  $K_c (A,Z)$ of a cluster of mass (charge) number $A$ ($Z$), are defined from the Gulderg and Waage mass action law in terms of the densities of free protons $\rho_p$ and neutrons $\rho_n$ as : 
\begin{eqnarray}
K_c (A,Z)=\frac{\rho_{AZ}}{\rho_p^{Z} \rho_n^{A-Z}} \, . \label{eq:chemical}
\end{eqnarray}
Here, $\rho_{AZ}$, $\rho_p$ and $\rho_n$ are, respectively,  the density of particles of the given species $(A,Z)$, the density of free protons and the density of free neutrons in a well defined thermodynamic condition, as given by the temperature $T$, total baryon density $\rho$ and proton fraction $y_p$. An experimental measurement of such constants requires the detection of particles and clusters from an equilibrated source, and an estimation of the associated thermodynamic parameters $(T,\rho,y_p)$. 

The experiment and data selection and processing were described in details in Ref.~\cite{BougaultJPG19}. Here we only recall the main features of the analysis.
Central  $^{136,124}$Xe+$^{124,112}$Sn collisions at 32 MeV/nucleon were detected with the INDRA apparatus, and a intermediate velocity source was selected using standard tools based on kinematical cuts. The detailed comparison between the four different reactions corresponding to the same bombarding energy per nucleon and different isospin ratios allowed to verify the statistical character of the emission, showing that for that sub-set of data chemical equilibrium is verified to  good approximation \cite{BougaultPRC97}.

Molecular dynamics calculations show that the expansion velocity of the moving source is monotonically decreasing with increasing emission time \cite{WangPRC72}.
Following Ref.~\cite{QinPRL108}, the time evolution of the expanding intermediate velocity source was therefore followed using the Coulomb corrected velocity of the particles in the source frame ($v_{surf}$) as a sorting variable.  
Increasing velocity corresponds to decreasing emission time.
 Under the assumption that chemical equilibrium holds at the different time steps of the emission, bins of $v_{surf}$ then correspond to statistical ensembles of particles in different thermodynamic conditions. 
The hypothesis of chemical equilibrium as a function of time was supported in Ref.~\cite{BougaultJPG19} by the observation that the extracted thermodynamic parameters as a function of $v_{surf}$ are independent of the entrance channel of the reaction.

The basic experimental observable in each $v_{surf}$ bin is given by the detected multiplicities $Y_{AZ}(v_{surf})\equiv \Delta N_{AZ}=\Delta v_{surf} dN_{AZ}/dv_{surf}$ of the different nuclear species, where $\Delta v_{surf}=0.2$ cm/ns  is the velocity bin. The measured particle numbers detected in a limited angular range $\Omega=\pi$ are normalized to the total solid angle assuming an isotropic emission.  

According to the expansion source picture that we are employing, the particles in the different $v_{surf}$ bins can be considered as a snapshot of the chemical distribution at the different times $t=t(v_{surf})$.
According to this picture, which is the crucial hypothesis of the analysis, the cluster densities at time $t$ can be estimated from the particle spectra  in the different bins of $v_{surf}$ as:

\begin{equation}
\rho_{AZ}=\rho\frac{\omega_{AZ}}{A}=\rho\frac{Y_{AZ}(v_{surf})}{\sum_{A,Z}AY_{AZ}(v_{surf})} \; ,
\end{equation}
where $\omega_{AZ}$ is the mass fraction of the cluster $(A,Z)$. The baryonic density at time $t$ is given by:
\begin{equation}
\rho=\frac{A_T}{V_T} \; , \label{eq:rho}
\end{equation}
where  $V_T$ is the effective volume of the expanding source at the emission time associated to the velocity bin under study, and $A_T$ is the corresponding total baryon number.
This latter is given by the measured multiplicities as:
\begin{equation}
A_T(t)= A_T(t=0)-\sum_{v_{surf}'>v_{surf}}  \sum_{A,Z} A Y_{AZ}(v_{surf}')  \, ,
\label{eq:atot}
\end{equation}
where the sum runs over the different velocity bins. $A_T(t=0)$ is the total baryon number
at the initial time of the expanding phase. This initial baryon number is calculated using a fitting procedure of the particle energy spectra. Neutrons are not detected in the experiment, and the free neutron number is deduced from the free proton number and the $^3$H/$^3$He ratio, assuming chemical equilibrium, as explained in \cite{BougaultJPG19}, and later in the text.

In this equation, we have considered that particles at velocity $v>v_{surf}$  are emitted earlier 
with respect to the time $t$ identified by the chosen velocity bin $v_{surf}$, and, therefore,
do not contribute to the corresponding statistical ensemble.

 The chemical constants can be expressed in terms of the mass fractions as:
\begin{eqnarray}
K_c (A,Z)=\frac{\omega_{AZ}}{A\omega_{11}^{Z} \omega_{10}^{A-Z}}\left( \frac{V_T}{A_T}\right )^{A-1} \, , \label{eq:chemica_final}
\end{eqnarray}
where $\omega_{AZ}$, $\omega_{11}$ and $\omega_{10}$ are the mass fractions of cluster (A,Z), proton and neutron respectively.

The measurement of the equilibrium constants thus requires an estimation of the source volume, at the different times of  the expansion. In previous studies \cite{QinPRL108,BougaultJPG19}, this was done with the Mekjian strategy often used in heavy ion collision analyses \cite{DasGuptaPR72}. Supposing an ideal gas of classical clusters in thermodynamic equilibrium at temperature $T$ in the grand-canonical ensemble, the differential spectrum of a cluster $(A,Z)$ is given by:
\begin{equation}
\frac{d^3N_{AZ}}{dp^3}(\vec p)=\frac{V_f}{h^3}  g_{AZ} \exp  \left [ \frac {1}{T} \left ( - \frac{p^2}{2M_{AZ}} + 
Z\mu_p + (A-Z)\mu_n \right )\right ] \; , \label{eq:spectra}
\end{equation}
where  $M_{AZ}=Am-B_{AZ}$ is the mass of cluster $(A,Z)$, $m$ is the nucleon mass, $\mu_n$ and $\mu_p$ are chemical potentials, $V_f$ is the free volume,  and the internal partition sum reads:
\begin{equation}
g_{AZ}=\sum_K (2J_K+1)\exp\left [ -\frac{E_{K}}{T}\right ]
\; ,
\end{equation}
where the sum runs over the different eigenstates of the clusters with energy $E_K$ and angular momentum $J_K$.
In the experimental sample, the differential spectra (corrected  for  the Coulomb boost as  
$\Delta p_A=\sqrt{2mA(E-ZE_C)}=Am \Delta v_{surf}$, with $E_C=10$ MeV \cite{BougaultJPG19}), are linked to the multiplicities by:
\begin{equation}
\frac{d^3N_{AZ}}{dp^3}(\vec p_A)\equiv\tilde Y_{AZ}= \frac{Y_{AZ}(v_{surf})}{4\pi p_A^2 \Delta p_A} \; . \label{eq:dN_Y}
\end{equation}

Assuming that Eq.~(\ref{eq:spectra}) holds, if we normalize the cluster spectrum  by the proton and neutron spectra at the same velocity, the unknown chemical potentials cancel and we get: 
\begin{equation}
\frac{\tilde Y_{AZ}(\vec {p_A})}{\tilde Y_{p}^Z(\vec p)\tilde Y_{n}^{A-Z}(\vec p)}=h^{3(A-1)}
\frac{2J_{AZ}+1}{2^A}\frac{1}{V_f^{A-1}} \exp \left [ \frac{B_{AZ}}{T}\right ] \; , \label{eq:ratio}
\end{equation}
where  $p_A=A\,p$, and $J_{AZ}, B_{AZ}$ are the angular momentum and binding energy of the ground state of the cluster, and we have neglected  the population of excited states. 

Eq.~(\ref{eq:ratio}) allows  independent estimations of  the free volume from the different cluster spectra as:
\begin{equation}
{ V_f}
=h^3 \exp \left [ \frac{B_{AZ}}{T(A-1)}\right ]\left (\frac {2J_{AZ}+1}{2^A}\frac{\tilde Y_{11}^Z(\vec p)\tilde Y_{10}^{A-Z}(\vec p)}
{\tilde Y_{AZ}(\vec p_A)} \right )^{\frac{1}{A-1}} \; . \label{eq:vf}
\end{equation}

In the absence of a direct neutron measurement, we can estimate the free neutron-proton ratio 
$R_{np}$ from the multiplicities of the $A=3$ isobars $^3$H and  $^3$He using Eq. (\ref{eq:spectra}):
\begin{equation}
R_{np}=\frac{\omega_{10}}{\omega_{11}}=\left(\frac{Y_{31}}{Y_{32}}\right)\exp{\left[ \frac {B_{32} - B_{31}}{T}\right]} \, , \label{eq:rnp}
\end{equation}
and Eq. (\ref{eq:vf}) becomes 
\begin{equation}
{ V_f}
=h^3R_{np}^{\frac{A-Z}{A-1}} \exp \left [ \frac{B_{AZ}}{T(A-1)}\right ]\left (\frac {2J_{AZ}+1}{2^A}\frac{\tilde Y_{11}^A(\vec p)}
{\tilde Y_{AZ}(\vec p_A)} \right )^{\frac{1}{A-1}} \; . \label{eq:vf1}
\end{equation}
Finally, the total volume entering  Eq.~(\ref{eq:chemica_final}) is computed from the free volume, given by Eq.~(\ref{eq:vf1}), as:
\begin{eqnarray}
V_T= V_f + \sum_{AZ}V_{AZ}\frac{\omega_{AZ}A_T}{A} \, ,
\label{eq:vt}
\end{eqnarray}
where we have  added the proper volume of the clusters which belong to the
source at a given time, with  $V_{AZ}=4\pi R_{AZ}^3/3$. $R_{AZ}$ is
the experimental radius of each cluster, taken from
Ref.~\cite{Angeli2013}.  In Sec.~\ref{sec:radius}, we  will
discuss how the choice of  $ R_{AZ}$ affects the determination of the
total volume, and, as a consequence, the total baryonic density and
the cluster densities.

It is important to stress that, for the consistency of the analysis, the different estimations of $V_f$ that can be obtained from Eq.(\ref{eq:vf1}) considering different particle species, that is different $(A,Z)$ values, should all coincide within experimental errors.
If this is not true as we will show in the following, the validity of either Eq.(\ref{eq:vf1}) or of the hypothesis of statistical equilibrium should be questioned.

To complete the analysis, the thermodynamic parameters $(T,y_p)$ must be evaluated together with the baryonic  density $\rho$, as a function of the surface velocity. 
The total  global proton fraction of the system is given by
\begin{eqnarray}
y_{p}=\frac{\sum_{A,Z} Z  Y_{AZ}}{\sum_{A,Z} A Y_{AZ}} \, ,\label{eq:yp}
\end{eqnarray}
with the neutron yield estimate from the proton yield via Eq.(\ref{eq:rnp}).
The temperature is obtained from the Albergo double ratio formula \cite{AlbergoNCA89}, that we briefly recall.
Integrating Eq.~(\ref{eq:spectra}) over momentum, and neglecting the excited states, we get:
\begin{equation}
N_{AZ}=V_f \left(\frac{M_{AZ}T}{2\pi\hbar^2}\right)^{3/2} (2J_{AZ}+1)e^{  \frac {  
Z\mu_p + (A-Z)\mu_n + B_{AZ} }{T}}\; . \label{eq:mult}
\end{equation}

The volume and chemical potential dependence can be eliminated by taking isobaric double ratios  \cite{AlbergoNCA89}.
In particular, using $^2$H, $^3$H, $^3$He and $^4$He, we have:
\begin{eqnarray}
T&=&\frac{\Delta B}{\ln\left [\frac{3}{2}\sqrt{\frac{9}{8}}R_v \right ] } 
\; ,
 \label{eq:temperature}
\end{eqnarray}
with 
\begin{eqnarray}
R_v&=&\frac{N_{21}N_{42}}{N_{32}N_{31}}= \frac{Y_{21}Y_{42}}{Y_{32}Y_{31}} \, ,\\
\Delta B&=& B_{21}+B_{42}-B_{32}-B_{31} \, .
\end{eqnarray}
We notice that a slightly different formula was used in previous analyses \cite{QinPRL108,HagelPRC62,KowalskiPRC75} with an extra $\sqrt{9/8}$ in the argument of the logarithm appearing in Eq.~(\ref{eq:temperature}) ($3/2\sqrt{9/8}\sqrt{9/8}$=$1.59\sqrt{9/8}$ instead of $3/2\sqrt{9/8}$). The justification of this extra factor was the fact that the analysis does not use total multiplicities, but particles at a given velocity described by a Maxwellian spectrum. We believe that this argument is not correct, because  the multiplicities  in a given $v_{surf}$ bin are proportional to the total particle numbers (see Eq.~(\ref{eq:mult})) and not to differential spectra  (see Eq.~(\ref{eq:dN_Y})).
However, we have verified that the presence (or absence) of an extra  $\sqrt{9/8}$ factor does not produce any visible effect in all the results shown in this paper.

Chemical constants measured by the NIMROD collaboration \cite{QinPRL108}, using the method explained above, were compared to statistical models employing different treatments of the nucleon-nucleon and nucleon-cluster interactions \cite{HempelPRC91,PaisPRC97}, with the purpose of constraining the expected cluster binding energy shifts in dense matter, and the associated Mott densities \cite{Roepke2015,RopkeNPA867,TypelPRC81}. 
However, it was observed in Ref.~\cite{BougaultJPG19} that the validity of the ideal gas expression, Eq.~(\ref{eq:spectra}), has to be assumed to determine  the thermodynamic state $(\rho,T,y_p)$, and also the value itself of the chemical constants via the free volume definition Eq.~(\ref{eq:vf1}). 
Indeed Eq.~(\ref{eq:spectra}) explicitly assumes that the cluster abundances are uniquely governed by their vacuum properties, notably their vacuum binding energies $B_{AZ}$, which is in contradiction with the very purpose of the analysis.
Moreover,  if in medium corrections were indeed negligible, the measured chemical constants would agree with the ideal gas prediction.
This latter can be easily worked  out from  Eq.~(\ref{eq:chemical}) considering that, for an ideal gas of clusters, $\rho_{AZ}=N_{AZ}/V_f$, with $N_{AZ}$ given by Eq.~(\ref{eq:mult}):
\begin{eqnarray}
K_c^{id}(A,Z)=\left(\frac{2\pi \hbar^2}{T}\right)^{\frac{3(A-1)}{2}}\left(\frac{M_{AZ}}{m^A}\right)^{3/2}\frac{(2J_{AZ}+1)}{2^A}  
\exp{\left[ \frac {B_{AZ}}{T}\right]} \, . \label{eq:ideal}
\end{eqnarray}

  \begin{figure}
  \begin{tabular}{c}
\includegraphics[width=1\textwidth]{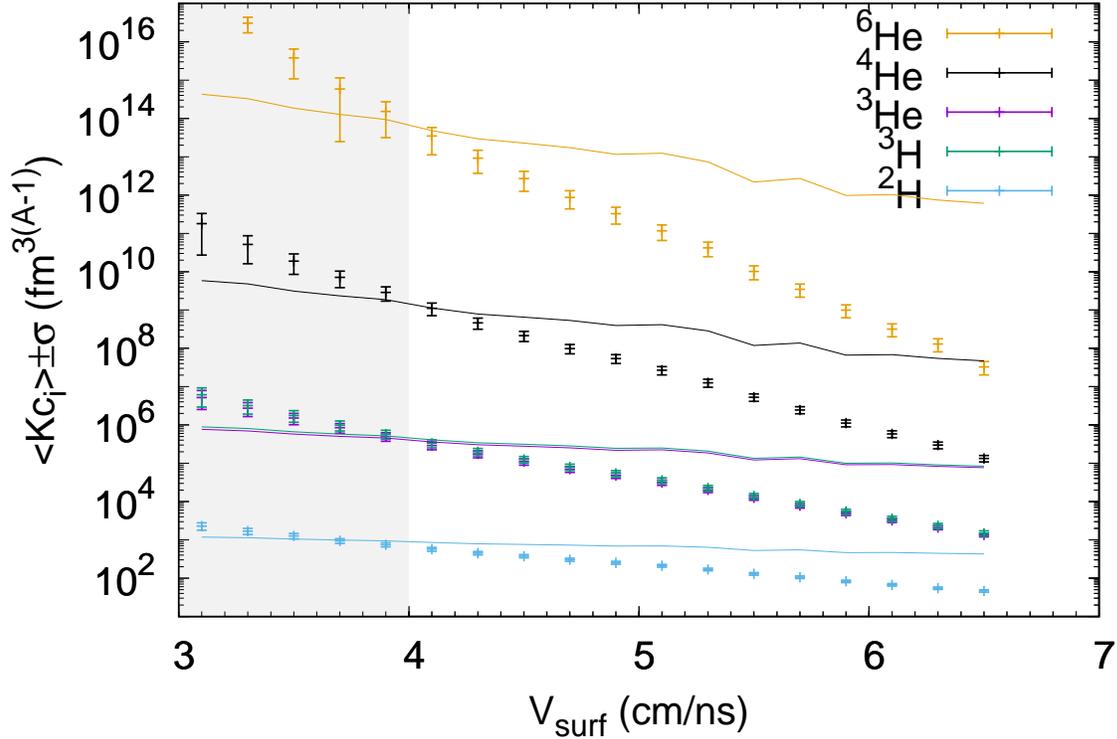}
  \end{tabular}
  \vspace{-1.cm}
\caption{(Color online) $^{124}$Xe$+^{124}$Sn system: The chemical equilibrium
  constants of each cluster  as a function of the surface velocity
  $v_{surf}$. The error bars only include statistical and systematic
  experimental errors, see Ref.~\cite{BougaultJPG19}. The solid lines
  are the ideal gas limit given by Eq.~(\ref{eq:ideal}). The grey band
  shows the area where data might be contaminated by emission from the spectator source.  }
\label{fig:start}
\end{figure}  

Chemical constants obtained from the analysis of the $^{124}$Xe$+^{124}$Sn system are displayed in
Fig.~\ref{fig:start}. 
In this figure, the value of the free volume used to estimate the baryonic density Eq.(\ref{eq:rho}) in each $v_{surf}$ bin is given by the arithmetic average of the volumes $ V_{f}^{(AZ)}$ extracted from Eq.(\ref{eq:vf1}) using the yields $Y_{AZ}$ of the different particle species :
\begin{equation}
\bar V_f=\frac{1}{N} \sum_{AZ} V_{f}^{(AZ)} \, , \label{eq:vaverage}
\end{equation}
with $N$ the total number of cluster species ($N=5$ in the present analysis).  In fact, this approach to define the system volume is largely arbitrary, since the different volume estimations from the different particle species are not compatible. In the following, a different approach will be considered for the determination of this key quantity, resulting in smaller volumes and having a two-fold effect on the equilirium constants, i.e.  making them  smaller and  moving them to a higher value of the density.

The only differences with respect to the results
published in Ref.~\cite{BougaultJPG19} are the inclusion of  the volume associated to deuterons in the arithmetic average of the present calculation and the normalization. Indeed it has to be noticed that the definition of chemical constants in Ref.~\cite{BougaultJPG19} differs  by a factor $A$ with respect to the one of Ref.~\cite{QinPRL108}. To allow an easier comparison with previous works, we have adopted the definition of Ref.~\cite{QinPRL108} in this paper.
As already mentioned above, the slightly different
expression for  the temperature Eq.~(\ref{eq:temperature})
does not produce any effect on the scale of the figure.  The
error bars are due to the experimental errors associated with the
measurements.   The solid lines represent the ideal gas limit, given
by Eq.~(\ref{eq:ideal}). The grey band shows the range where the
experimental data might be contaminated by emission from the spectator source, since the proton spectra are
not well reproduced by the fit used to deduce the mass of the evolving source. 
One should also observe that equilibrium constants of $^3$H and $^3$He are so close, that they are almost indistinguishable on the scale of the figure.
We can see that the measured chemical constants are systematically lower than the ideal gas prediction, and the effect increases with increasing density, showing that binding energy shifts are necessary.
A qualitatively similar deviation from the ideal gas limit was also found in NIMROD data \cite{QinPRL108,HempelPRC91}. 
It is, therefore, clear that a correction is needed to Eq.~(\ref{eq:spectra}) for the analysis to be consistent. If in-medium corrections at a given temperature and density only depend on the baryonic number of the particle, then their effect will cancel out when taking isobaric ratios and double ratios as in Eqs.(\ref{eq:rnp}) and (\ref{eq:temperature}). However, this is not the case for the volume $V_f$, Eq.~(\ref{eq:vf1}), which in turn affects both the evaluation of the densities $\rho_{AZ}$ and the evaluation of the total baryonic density $\rho$.

\begin{figure}
  \begin{tabular}{cc}
\includegraphics[width=1\textwidth]{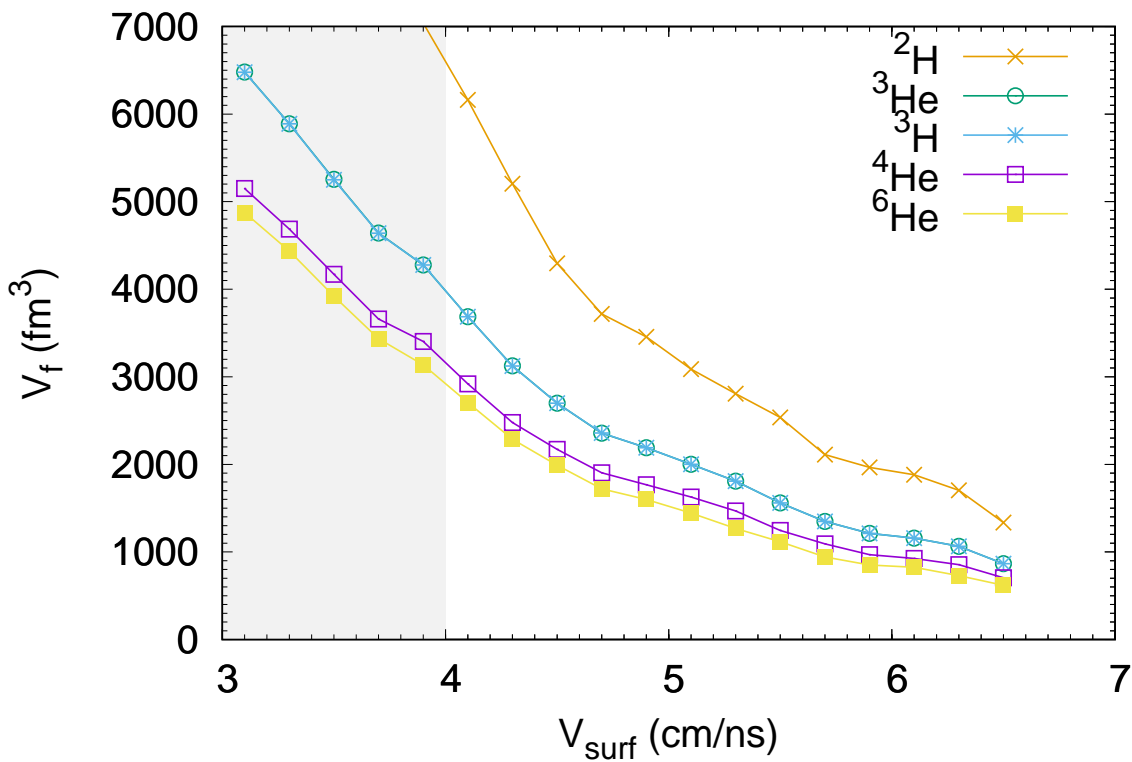}
  \end{tabular}
  \vspace{-1cm}
\caption{(Color online) $^{124}$Xe$+^{124}$Sn system: The free volume, as estimated from the yields of different cluster species, as  a function of $v_{surf}$. Results for $^3$H and $^3$He are indistinguishable. The grey band shows  the area where data  might be contaminated  by emission from the spectator source.  }
\label{fig2}
\end{figure}

The need of an in-medium correction to the ideal gas expression  Eq.~(\ref{eq:vf1}) is further shown by Fig.~\ref{fig2}.
This figure displays the value of the free volume obtained from  Eq.~(\ref{eq:vf1}) as a function of the sorting variable $v_{surf}$ for the $^{124}$Xe$+^{124}$Sn system, using different particle species. Since a $v_{surf}$ bin represents a specific thermodynamic condition $(\rho,T,y_p)$,  if everything was consistent, we should find the same volume whatever the cluster species considered, which is clearly not the case except for the $A=3$ isobars, which lead to almost identical volume estimations (indistinguishable on the scale of the figure). 
Qualitatively similar results were obtained with the NIMROD data \cite{QinPRL108}, showing  that 
the incompatibility among the different volume estimations is not an experimental problem, but it rather points towards  an inconsistency in the analysis method. 
To solve this inconsistency,  in the next section we introduce a modification in  Eq.~(\ref{eq:spectra}) allowing for possible in-medium effects.

\section{Bayesian analysis} \label{sec:analysis}

Since the different in-medium effects that contribute to the determination of cluster multiplicities can be seen as a shift of the cluster binding energy  \cite{Roepke2015,RopkeNPA867,TypelPRC81}, it is reasonable to suppose that the correction can be defined as a Boltzmann factor:

\begin{eqnarray}
C_{AZ}&=&\exp \left [- \frac{\Delta_{AZ}}{T(A-1)}\right ] \label{eq:correction} \, ,
\end{eqnarray}
with $\Delta_{AZ}$ given by
\begin{eqnarray} \label{eq:DeltaAZ}
\Delta_{AZ}=a_1 A^{a_2}+a_3 |I|^{a_4} \, ,
\end{eqnarray}
where $a_i$, $i=1,\dots,4$ are free parameters.
The dependence on the cluster species is given by the term $\Delta_{AZ}$, which, 
  for each thermodynamic condition $(T,\rho,y_p)$  identified by a bin in $v_{surf}$, 
 can in principle  depend  on the two good quantum numbers of each cluster, $A$ and $I=(2Z-A)/2$. 
The influence of the functional expression for the correction $\Delta_{AZ}$ is studied in Section \ref{sec:correction}. 

It is important to stress that $\Delta_{AZ}$ can be interpreted as a reduction of the binding energy only in the framework of the classical simplified ideal expression 
Eq.(\ref{eq:spectra}). Indeed the presence of a nuclear medium affects nucleons both in bound, unbound and resonant states \cite{TypelPRC81}, and $\Delta_{AZ}$ should be 
understood as a global effective correction accounting for all the missing quantum and interaction effects, such as Pauli blocking, effective masses, couplings to the mesons, etc.
This correction modifies the expression of the free volume $V_f$ as it can be estimated from the abundance of a given $(AZ)$ species:
\begin{equation}
V_f=h^3R_{np}^{\frac{A-Z}{A-1}} C_{AZ} \exp \left [ \frac{B_{AZ}}{T(A-1)}\right ] 
\cdot \left (\frac {2J_{AZ}+1}{2^A}\frac{\tilde Y_{11}^A(\vec p)}
{\tilde Y_{AZ}(\vec p_A)} \right )^{\frac{1}{A-1}} \; , \label{eq:vfnew}
\end{equation}
where the temperature $T$ is still estimated by eq.(\ref{eq:temperature}).
The unknown  parameters $\vec a =\{ a_i(\rho,y_p,T),i=1 - 4\}$ can be fixed by imposing that the volumes obtained from the experimental spectra $\tilde Y_{AZ}$ via Eq.(\ref{eq:vfnew}) of the different $(A,Z)$ nuclear species in a given $v_{surf}$, correspond to compatible values. Because of the presence of experimental uncertainties, we cannot simply solve Eq.~(\ref{eq:vfnew})  for the $\vec a$ parameters to impose a strictly identical volume for the different species. Even if the experimental errors were negligible, 
the correlation between $v_{surf}$ and the volume is not a one-to-one correlation because of the physical dispersion of the $v_{surf}$ variable. For these reasons, we consider the unknown $\vec a$ parameters as random variables. We take in each $v_{surf}$ bin flat priors,  $P_{\rm prior}(\vec a)=\theta(\vec a_{\rm min}-\vec a_{\rm max})$, within an interval largely covering the physically possible reduction range of the  binding energy, $0\le a_1\le 15$ MeV, $0\le a_3\le a_1$ MeV, $-1\le a_2 \le 1$, $0 \le a_4 \le 4$.  

The posterior  distribution is  obtained by imposing the volume observation with a likelihood probability as follows:
\begin{eqnarray}
P_{post}(\vec a)={\cal N}\exp\left(-\frac{\sum_{AZ}(V_{f}^{(AZ)}(\vec a)-\bar V_f(\vec a))^2}{2\bar V_f(\vec a)^2}\right) \, . \label{eq:likely}
\end{eqnarray}  
Here, ${\cal N}$ is a normalization, $V_{f}^{(AZ)}(\vec a)$ is the free volume obtained from the $(A,Z)$ cluster using Eq.(\ref{eq:vfnew})  with the specific choice $\vec a$ for the parameter set of the correction, and $\bar V_f(\vec a)$ is the average volume corresponding to a given parameter set $\vec a$ from Eq.(\ref{eq:vaverage}).

 \begin{figure*}
  \begin{tabular}{cc}
    \includegraphics[width=0.5\textwidth]{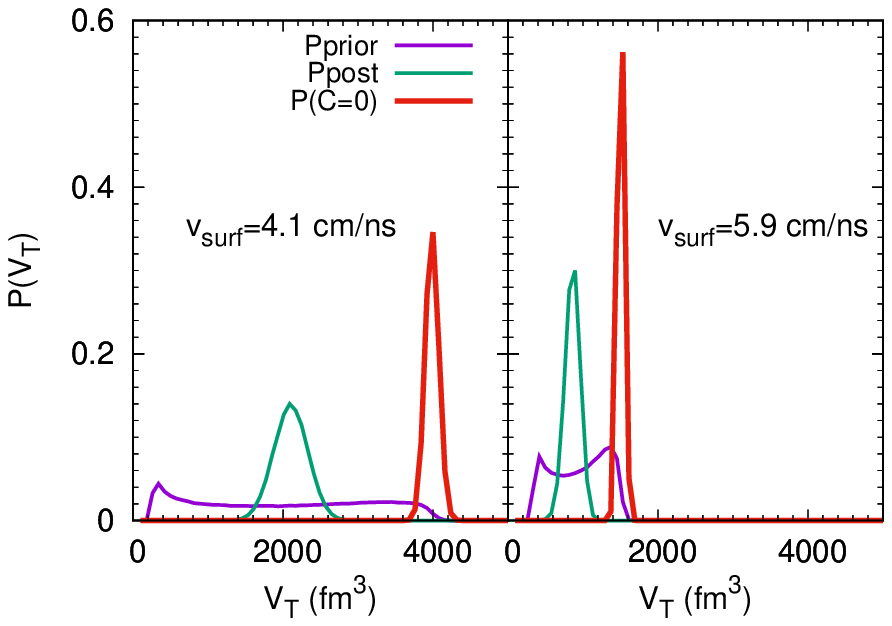} &  \includegraphics[width=0.5\textwidth]{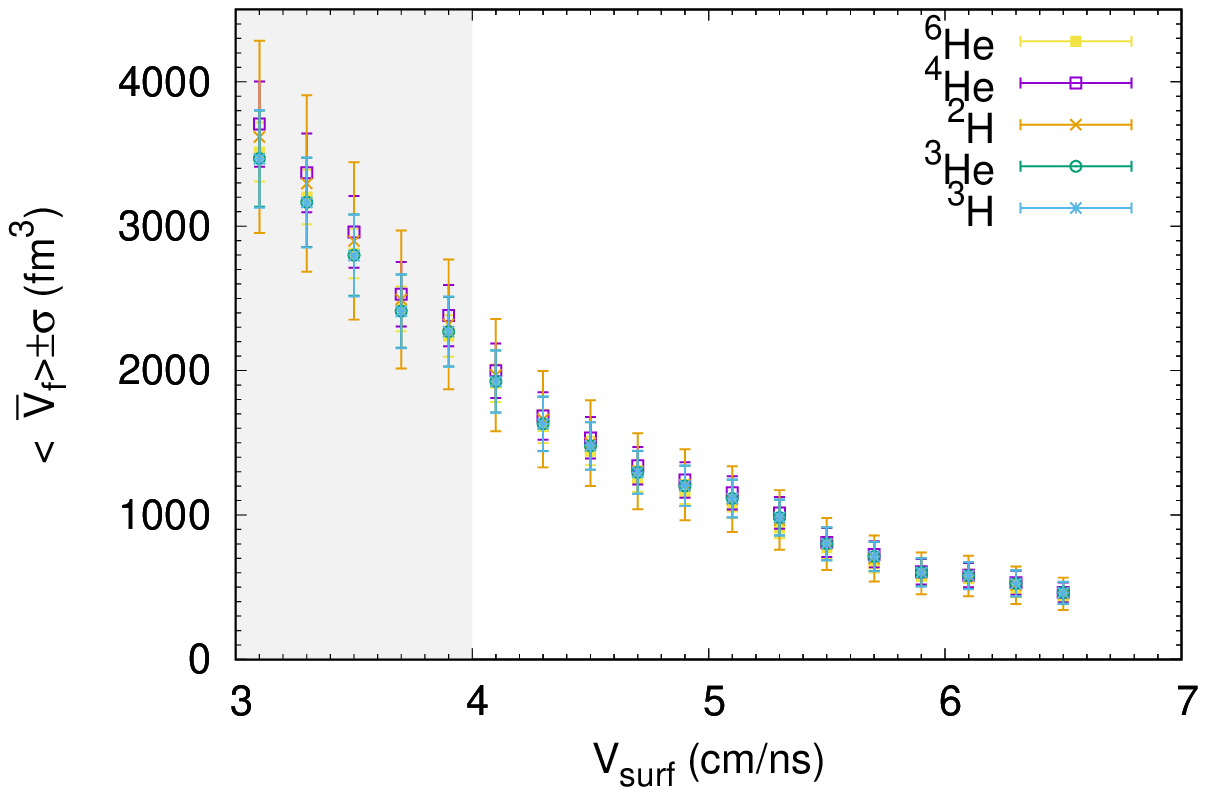}
  \end{tabular}
\caption{$^{124}$Xe$+^{124}$Sn system: Left: The prior (purple) and
  posterior (green) probability distributions of the total volume for the bins with $v_{surf}=4.1$ cm/ns (left) and $v_{surf}=5.9$ cm/ns (right). In red, the probabilities without the correction, designated by $P(C=0)$, are also  shown. Right: The corrected free volumes associated with the clusters as a function of
  $v_{surf}$. The error bars are due to both the correction and the experimental errors. The grey band shows the area where data might be contaminated by emission from the spectator source. }
\label{fig3}
\end{figure*}

The prior (posterior) probability distribution of any physical quantity $X$ is then readily calculated as:
\begin{equation}
P(X=X_0)=\int d\vec a P(\vec a) \delta\left ( X(\vec a) - X_0 \right ) \, ,
\end{equation}
where $P(\vec a)$ is the prior (posterior) distribution of the correction parameters.
Similarly, expectation values can be calculated as:

\begin{eqnarray}
\langle X \rangle =\int d \vec a  P(\vec a) X(\vec a) \, ,
\end{eqnarray}
and the correspondent standard deviations as,
\begin{eqnarray}
\sigma_X =\sqrt{\langle X^2\rangle - \langle X\rangle^2 } \, .
\end{eqnarray}

The left part of Fig.~\ref{fig3} shows the prior and posterior distribution of the  total volume  $V_T$ in two chosen velocity bins, $v_{surf}=4.1$ cm$/$ns (6$^{\rm th}$ bin) and $v_{surf}=5.9$ cm$/$ns (15$^{\rm th}$ bin). In red, we also show the probabilities calculated without the correction factor, using Eq.(\ref{eq:vf1}). In this case, the width of the  distribution is only due to the experimental errors.
The general effect of allowing for an in-medium correction is a
reduction of the estimated volume. This can be immediately understood
by comparing the ideal gas estimation Eq.(\ref{eq:vf}), and the
modified expression Eq.(\ref{eq:vfnew}), and considering that we
impose that the correction $C_{AZ}\le 1$.
Indeed, following the microscopic calculations of in-medium effects \cite{Roepke2015}, we consider that, because of the Pauli blocking effect, the influence of the external nucleon gas goes in the direction of reducing the effective binding in the medium.

We can see that in the  lower velocity bin, corresponding to a later time and larger volume, the prior volume distribution is completely unconstrained, reflecting the large dispersion of the volume estimation which is obtained in the absence of the correction (see Fig.\ref{fig2}).
On the other hand, the condition of compatibility between the volume measurements allows a fair determination of this variable, crucial for the rest of the analysis. In the higher velocity bin, corresponding to earlier times of the expansion and more compact configurations, the volume scale is reduced, meaning that the volume estimation is less sensitive to the importance of the correction.
Still, the posterior distribution is considerably narrower than the prior one.  
The corrected expectation values of the volume as estimated from the multiplicities of each cluster from Eq.(\ref{eq:vfnew}), with the associated standard deviations, are shown as a function of $v_{surf}$  in the right part of  Fig.~\ref{fig3}. It is clear that when we include the correction, the average volumes  are systematically lower than the uncorrected results of Fig.\ref{fig2},
 and the estimations obtained from the different cluster species  are compatible within error bars.

\section{Equilibrium constants} \label{sec:kci}

\begin{figure}
  \begin{tabular}{cc}
\includegraphics[width=0.7\textwidth]{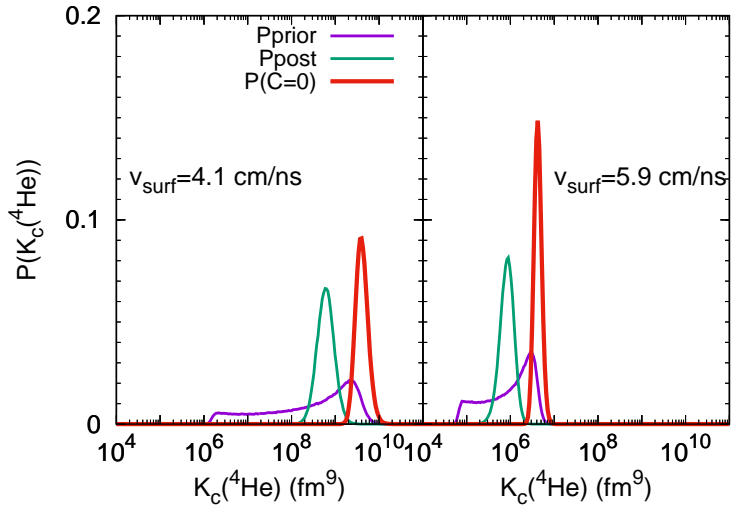} 
  \end{tabular}
 \caption{ The prior (purple) and posterior (green) probability distributions of the chemical equilibrium constant of the $\alpha$ cluster for the bins with $v_{surf}=4.1$ cm/ns (left) and $v_{surf}=5.9$ cm/ns (right) for the $^{124}$Xe$+^{124}$Sn system.  In red, the probabilities without the correction, $P(C=0)$, are also shown. }
\label{fig4}
\end{figure}

\begin{figure}
  \begin{tabular}{cc}
 \includegraphics[width=0.8\textwidth]{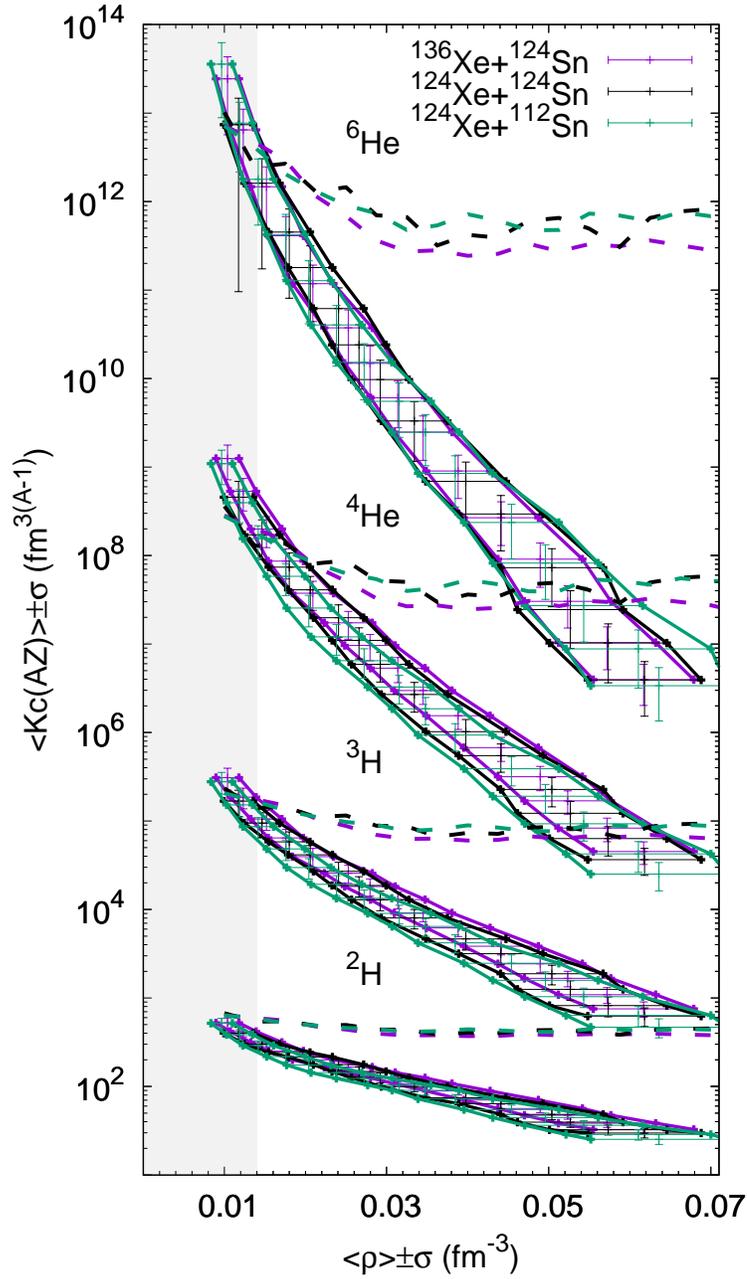}
  \end{tabular}
  \vspace*{-0.8cm}
\caption{(Color online)  The chemical equilibrium constant of all the clusters
  as a function of the density for the three experimental systems. 
    The error bars are due to the correction and the experimental
    errors. The dashed lines are the ideal gas limit given by Eq.~(\ref{eq:ideal}). The grey band shows the area where data might be contaminated  by emission from the spectator source.}
\label{fig5}
\end{figure}

 The difference between the volume estimation from the ideal gas
 assumption Eq.(\ref{eq:vf}), and the one determined by the more
 general expression  Eq.(\ref{eq:vfnew}) with the Bayesian estimation
 Eq.(\ref{eq:likely}) of the correction parameters $\vec a$, 
obviously reflects  itself on the estimation of the chemical constants, as shown in Figs.~\ref{fig4} and \ref{fig5}. 
Similar to Fig.\ref{fig3}, we show in Fig.\ref{fig4} the prior and posterior distribution of the $\alpha$ chemical constant  $K_c(4,2)$ in  the same velocity bins analyzed in Fig.\ref{fig3}, while Fig.~\ref{fig5} shows the average and variance of the chemical equilibrium constants for all the clusters as
a function of the density for the three different experimental systems.
Comparing the uncorrected (labelled "P(C=0)") and corrected (labelled "Ppost") distributions in Fig.\ref{fig4}, 
we can see that the correction to the ideal  gas hypothesis leads to a systematic decrease of the chemical  constants with respect to the ideal gas assumption (\ref{eq:vf1}), and an increased dispersion. The effect is higher at lower density, but a shift as important as a factor 10 is observed also in the highest velocity bin, corresponding to the highest density.

\begin{figure}
  \begin{tabular}{cc}
\includegraphics[width=0.9\textwidth]{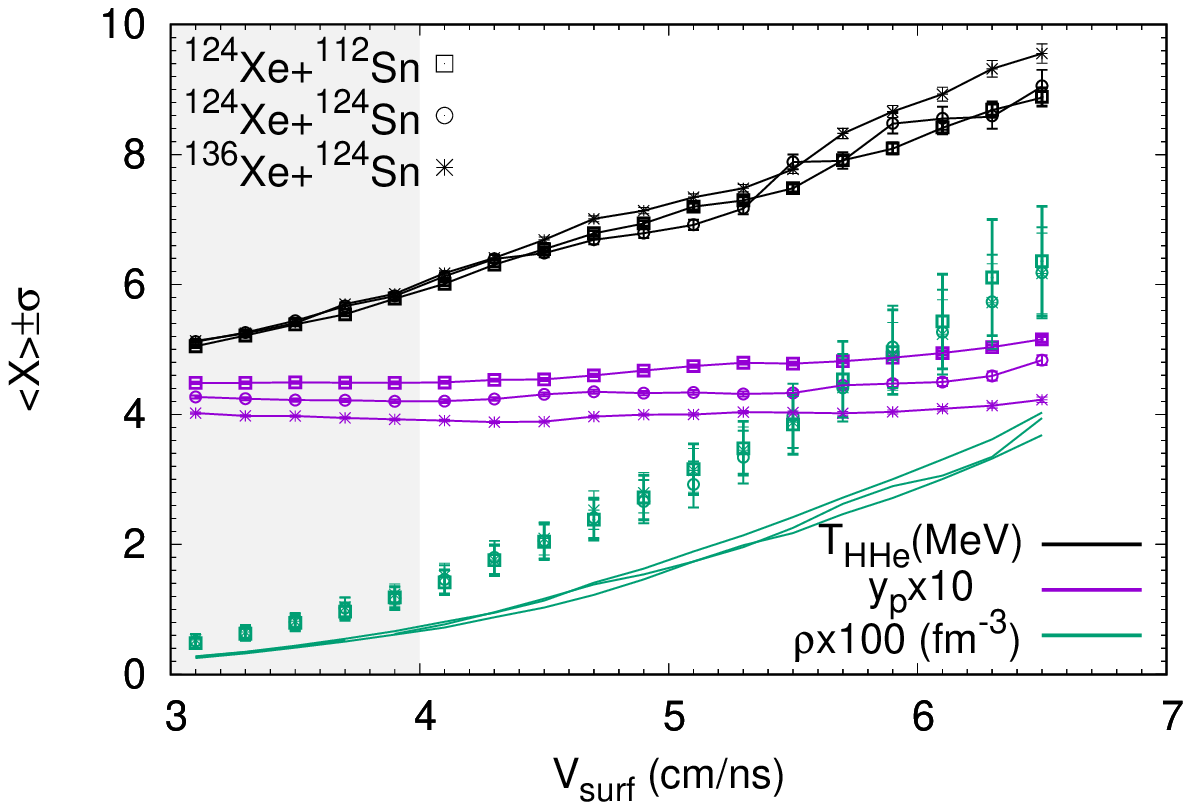}
  \end{tabular}
  \vspace*{-0.8cm}
\caption{Symbols: double isotope ratio temperature $T_{HHe}$ (black), global proton fraction (multiplied by a factor of 10) (purple),  baryon density (multiplied by a factor of 100) (green) as a function of the experimental quantity $v_{surf}$ for the three experimental systems. The uncertainties reflect both the correction and the experimental errors. The solid lines report the ideal gas limit, and were obtained from eq. (\ref {eq:vaverage}) with eq. (\ref{eq:vf1}). The grey band shows the area where
data might be contaminated by emission from the spectator source.} 
\label{fig:thermo}
\end{figure}

The chemical equilibrium constants of all the clusters measured
in the three different experimental data sets are displayed in Fig.\ref{fig5}. The dashed lines show the ideal gas limit, given by Eq.(\ref{eq:ideal}), which, just like in Fig.~\ref{fig:start}, give incompatible values with the experimental data. For the corrected case, we can see that the results of all the three systems almost perfectly overlap, confirming the expectation that chemical constants do not depend on the proton fraction of the system. It would be interesting to check this point with a larger proton-neutron asymmetry.

\begin{figure*}
  \begin{tabular}{cc}
\includegraphics[width=1.\textwidth]{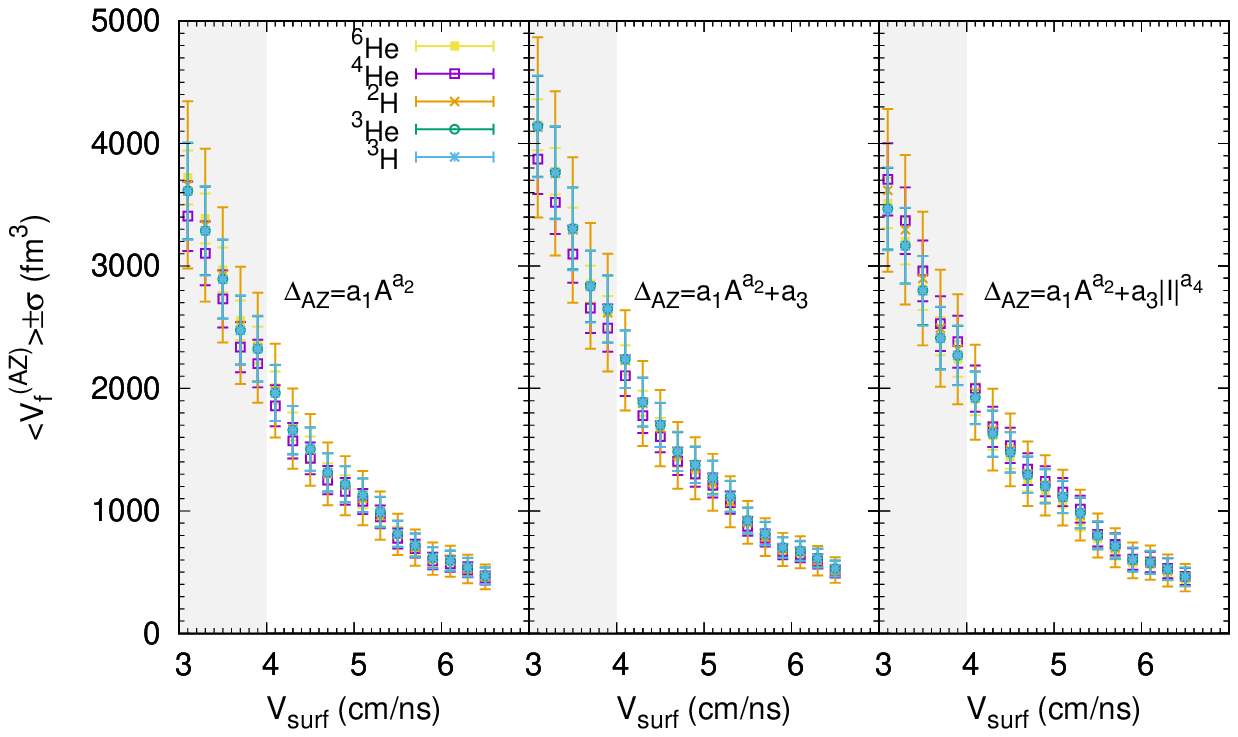}
  \end{tabular}
  \vspace*{-1.2cm}
\caption{(Color online) $^{124}$Xe$+^{124}$Sn system. Average volume estimated from the different 
  clusters and standard deviations as a function of  $v_{surf}$ for three different functional forms of the correction $\Delta_{AZ}$. The uncertainties are due to both the correction and the experimental errors. The grey band shows the area where data might be contaminated by emission from the spectator source.}
\label{fig:corrections-V}
\end{figure*}  

The thermodynamic conditions explored in the experiments are displayed in Fig.\ref{fig:thermo}. The temperature is here estimated through the double isotope formula Eq.(\ref{eq:temperature}), and is indicated as $T_{HHe}$.
We can see that the different collisions explore very similar trajectories in the $(T_{HHe},\rho)$ plane, the only difference being in the global proton fraction, as expected. 
Comparing with the results of Ref.~\cite{BougaultJPG19}, where the ideal gas equation Eq.(\ref{eq:vf1}) was used to estimate the volume following Ref.~\cite{QinPRL108}, (full lines in Fig.\ref{fig:thermo}), we can see that the existence of an in-medium correction goes in the direction of increasing the density, and the effect is the same for the three systems.

\begin{figure*}
  \begin{tabular}{cc}
\includegraphics[width=1.\textwidth]{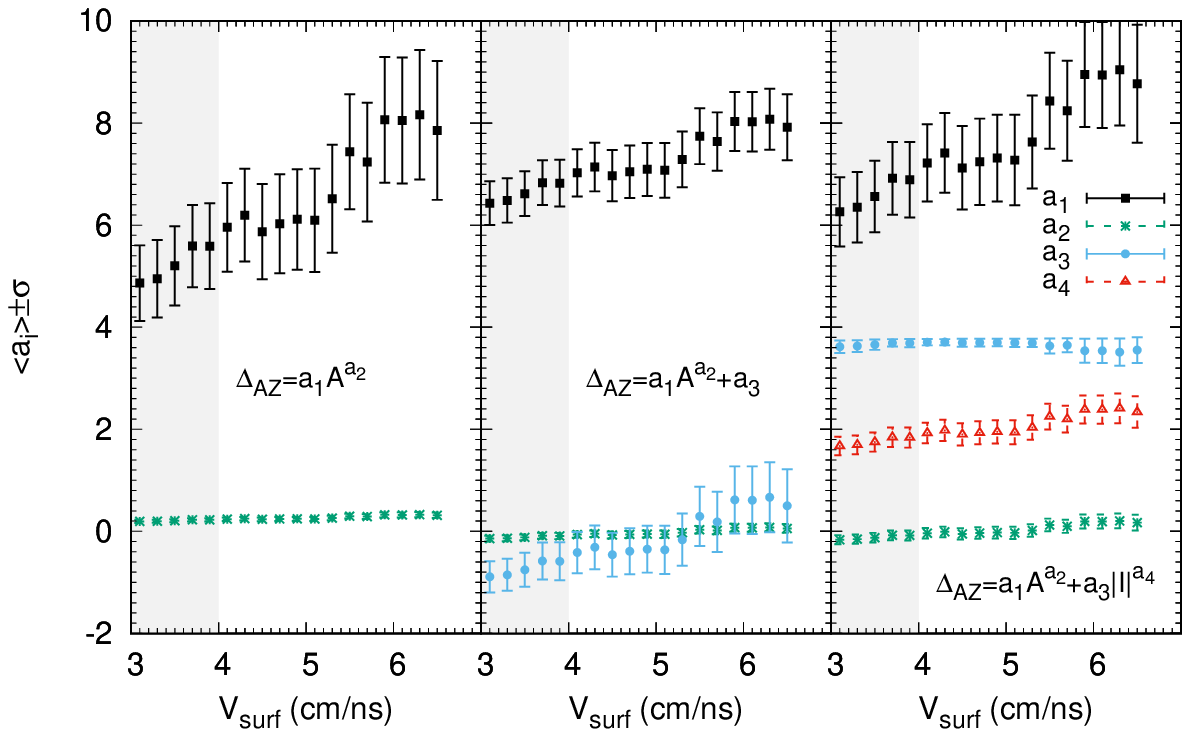}
  \end{tabular}
   \vspace*{-1.2cm}
\caption{(Color online) $^{124}$Xe$+^{124}$Sn system. Posterior estimation of the average and standard deviation of the parameters $a_i$  as a function of $v_{surf}$ for three different functional forms of the correction $\Delta_{AZ}$. The uncertainties are  due to both the correction and the experimental errors. The grey band shows the area where data might be contaminated by emission from the spectator source.}
\label{fig:corrections-ai}
\end{figure*}

Still, it is important to stress that the temperature is evaluated with the Albergo $T_{HHe}$ thermometer of Eq.(\ref{eq:temperature}). As we have discussed in Section \ref{sec:form}, this expression corresponds to the true thermodynamical temperature $T$ only if the in-medium corrections to the ideal gas of clusters expression Eq.~(\ref{eq:mult}) cancel 
in the double ratio. This is in principle not the case if the correction does not scale linearly with the particle numbers. 
Our Bayesian analysis does not allow us to determine the deviation of Eq.(\ref{eq:temperature}) from the true thermodynamic temperature, and this can only be done in the framework of a specific model.
One such model will be considered in Section \ref{sec:model}.

\begin{figure*}
  \begin{tabular}{cc}
\includegraphics[width=1.\textwidth]{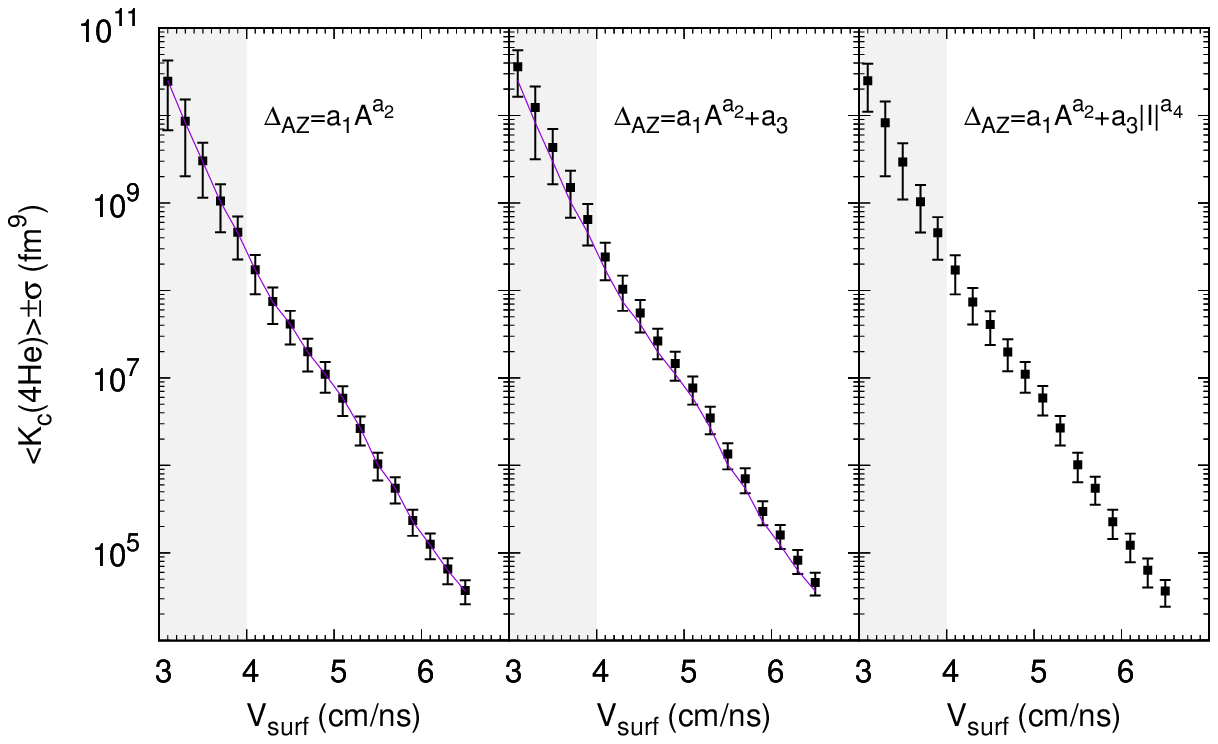}
  \end{tabular}
  \vspace*{-1.2cm}
\caption{(Color online) $^{124}$Xe$+^{124}$Sn system. $\alpha-$particle chemical equilibrium constants and standard deviations as a function of $v_{surf}$ for three different functional forms of the correction $\Delta_{AZ}$. The full lines in the left and middle panels correspond to the average right panel results. The uncertainties are due to both the correction and the experimental errors. The grey band shows the area where data might be contaminated by emission from the spectator source.}
\label{fig:corrections-Kc}
\end{figure*}

\subsection{Different parameter sets for the in-medium effect correction} \label{sec:correction}

The correction given by Eq.(\ref{eq:DeltaAZ}) is a four-parameter set, function of  the number of nucleons $A$  and isospin $I$. 
This functional form has a certain degree of arbitrariness, and the expression  is not unique. 
 In principle, the correction can depend on all the good quantum numbers of the clusters, namely $A$, $I$, and the charge $Q$. However, the volumes extracted from $^3$H and $^3$He are fully compatible already in the uncorrected data, meaning that, whatever the in-medium correction, it must be the same for both clusters. As a consequence, the correction cannot depend on $Q$. The compatibility of the volumes for $^3$H and $^3$He also implies that we have four data points  in each $v_{surf}$ bin  to fit our four parameters. This means that in practice we are extracting independent corrections for the different particle species. This is done on purpose, to insure that we are not introducing unjustified hypotheses on the functional dependence of the in-medium correction. However, the drawback is that we cannot attribute a clear physical meaning to the functional form we have introduced. To progress on this point, the analysis should be extended to other cluster species, to see if a four-parameter set is still enough to describe the whole set of data. In the absence of this extra experimental information, we can see if a reduced number of parameters is sufficient to describe the present data, and if 
 varying the number of parameters we can get different results for the equilibrium constants. To this aim, we have introduced three different parameterisations, employing two, three and four parameters, respectively,  
 and the results are reported in Figs.~\ref{fig:corrections-V}, \ref{fig:corrections-ai}, and \ref{fig:corrections-Kc}. Fig. ~\ref{fig:corrections-V} shows the posterior first (data point) and second (error bar) moment of the volume distribution. Fig.~\ref{fig:corrections-ai} shows the optimal values obtained for the parameters, and Fig.~\ref{fig:corrections-Kc} presents the posterior chemical equilibrium constant for the $\alpha-$particle. 
 
The analysis of Figs.~\ref{fig:corrections-V} and \ref{fig:corrections-Kc} reveals
  that, whatever the hypothesis on the functional form, we get fully compatible results for the average volumes and chemical constants, which implies compatible results for all observables. 
 In particular, in the left panels of Figs.~\ref{fig:corrections-V}, \ref{fig:corrections-ai}, and \ref{fig:corrections-Kc}, a simple two-parameters prescription  $\Delta_{AZ}=a_1 A^{a_2}$ is employed. We can see that the compatibility between the different volume estimations is comparable to the one which is obtained with the full four-parameters formula presented in the right panels. This means that we do not have a compelling evidence that the in-medium correction is isospin dependent.
The comparison between the extracted values of the parameters with the different parametrizations  is also instructive. Within the simplest prescription shown in the left panel of Fig.~\ref{fig:corrections-ai}, the value of the $a_2$ parameter is compatible with zero, meaning that the optimal solution is roughly compatible with a constant, that is an effective correction $\Delta_{AZ}$ that depends on the thermodynamic condition but does not depend on the particle species.
This is confirmed by the comparison between the middle panels of Figs.~\ref{fig:corrections-V}, \ref{fig:corrections-ai}, and \ref{fig:corrections-Kc}, where an isospin independent three-parameter formula $\Delta_{AZ}=a_1 A^{a_2}+a_3$ is tested. We can see that $a_2$ is still compatible with zero, and the sum $a_1+a_3$ is roughly equal to the $a_1$ value obtained in the simpler prescription. This means that the functional dependence on $A$ of the correction appears relatively robust, and confirms that an $A$ independent correction seems to be sufficient.
It is important to stress that this interesting experimental finding 
does not imply that  the in-medium binding energy shift of light clusters is really a constant; it only shows that the effect  does not have any clear monotonic dependence with the baryonic number in the range $A=2-6$. Indeed the posterior volume dispersion is never negligible (see Fig.\ref{fig3}), suggesting that a dependence on the nuclear species cannot be excluded. Beyond mean-field theoretical calculations are needed to settle this point \cite{Roepke2015}.

Finally, the right panels of Figs.\ref{fig:corrections-V}, \ref{fig:corrections-ai}, and \ref{fig:corrections-Kc} show the complete four-parameter formula $\Delta_{AZ}=a_1 A^{a_2}+a_3|I|^{a4}$ that will be used as fiducial expression in the successive figures. We can see that the independence on the cluster size is again confirmed, while the data point towards an (approximately quadratic) isospin dependence. As already mentioned above, this is not however a conclusive argument because we have as many parameters as particle species, and a confirmation of the possible isospin dependence of the in-medium effects would need extra equilibrium constants with different isospin values.

 \subsection{Effect of the radius of the clusters} \label{sec:radius}
 
As already observed from Fig.\ref{fig:thermo}, allowing for the possibility of in-medium corrections 
in each $v_{surf}$ bin, 
globally leads to smaller volumes (and therefore higher global baryonic densities), with respect to the previous works Refs.\cite{QinPRL108,BougaultJPG19}, where the hypothesis of an ideal gas of clusters was explicitly made in the data analysis. This density increase stems from the fact that we expect from basic theoretical arguments \cite{Roepke2015} that the in-medium corrections should reduce the effective binding. This corresponds to a positive $\Delta_{AZ}$ (see Eq.(\ref{eq:correction})), and consequently smaller volumes (see Eq.(\ref{eq:vfnew})). However, in order to quantitatively extract the density value, 
the   proper volume of the clusters has to be added to the free volume, see  Eq.(\ref{eq:vt}), which leads to an extra uncertainty in the analysis.

In this work we have chosen for the radius of each clusters the
 experimental values given in Ref. \cite{Angeli2013} but in previous
 works the authors have chosen a fixed value for the radius parameter of 1.3 fm. In this section we want to discuss the effect that this might
 have in both the thermodynamics and the chemical equilibrium
 constants. The temperature and
 proton fraction are not affected by the value of $r_0$, but, as seen In Fig.~\ref{fig:r0-kc4he}, there is a
 clear effect on the total baryonic density, with smaller values of
 $r_0$ giving larger densities.  Due to the influence of this
 parameter, we have decided to take in our data analysis
 the  experimental radius of each cluster $R_{AZ}$ taken from
Ref.~\cite{Angeli2013}, and the results are also shown in the same figure.
The use of the experimental value $R_{AZ}$ gives rise to
the smallest densities.
In Fig.~\ref{fig:r0-kc4he}, the chemical equilibrium constant for the
 $\alpha-$particle is plotted as a function of $v_{surf}$ (left panel) and total
 baryonic density (right panel), considering the total volume $V_T$ determined by
 $R_{AZ}$ and by  {fixed} $r_0$. For the lowest
 densities, all calculations coincide. At the other $v_{surf}$
 extreme, corresponding to the  largest densities,  $R_{AZ}$
 estimates smaller maximum densities, and, for a given density, smaller
 equilibrium constants.

   \begin{figure}[!htbp]
  \begin{tabular}{cc}
\includegraphics[width=0.7\textwidth]{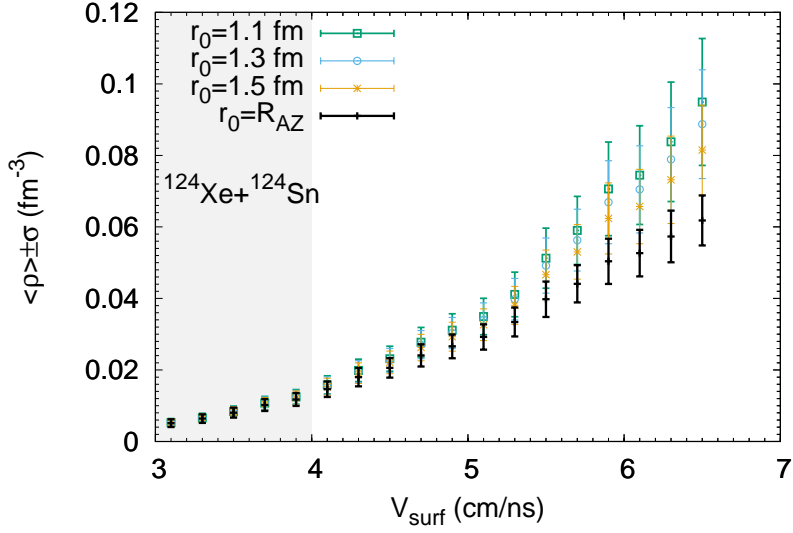}
  \end{tabular}
  \vspace*{-0.8cm}
\caption{(Color online) $^{124}$Xe$+^{124}$Sn system: The density as a
  function of the experimental quantity $v_{surf}$ for the four
  different radii considered. The uncertainties reflect both the
  correction and the experimental errors. The grey
band shows the area where data might be contaminated by emission from the spectator source.}
\label{fig:r0-rho-Vsurf}
\end{figure}

   \begin{figure}[!htbp]
  \begin{tabular}{cc}
\includegraphics[width=1\textwidth]{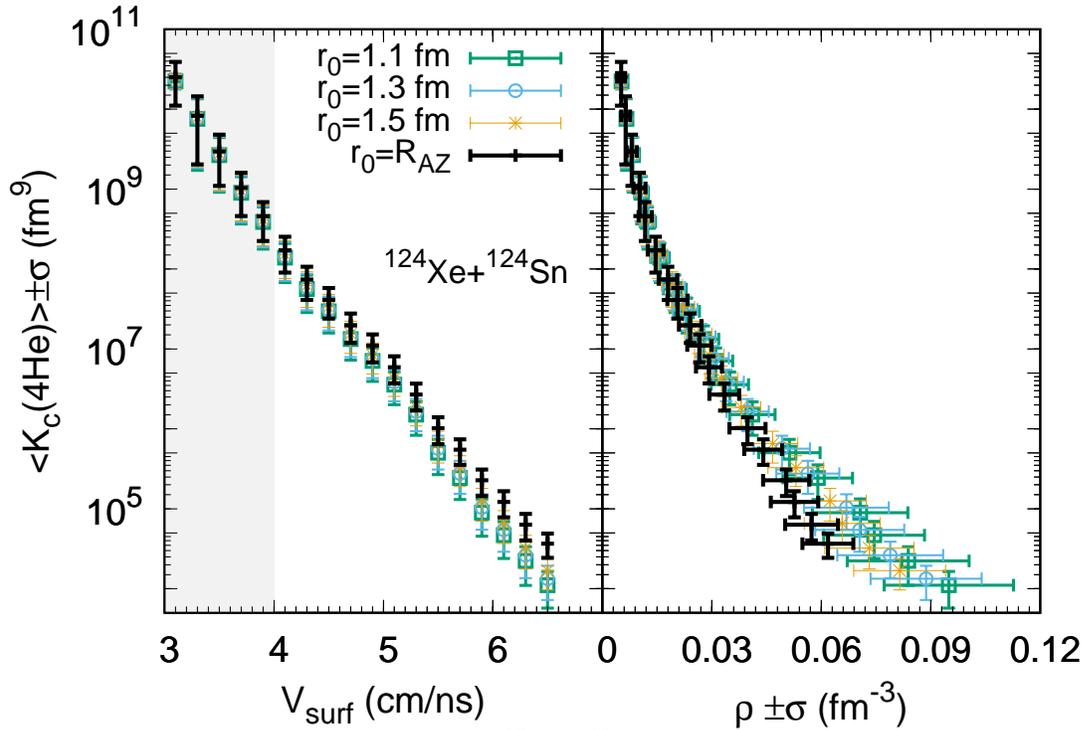}
  \end{tabular}
  \vspace*{-0.7cm}
\caption{(Color online) $^{124}$Xe$+^{124}$Sn system: The chemical equilibrium constant for
  the $\alpha-$particle as a function of $v_{surf}$ (left) and as a
  function of the density (right)  for the four
  different radii considered. The uncertainties reflect both the
  correction and the experimental errors. The grey
band shows the area where data might be contaminated by emission from the spectator source.}
\label{fig:r0-kc4he}
\end{figure} 

\section{Comparison with a theoretical model} \label{sec:model}

\begin{figure}
 \begin{tabular}{c}
 \includegraphics[width=0.8\textwidth]{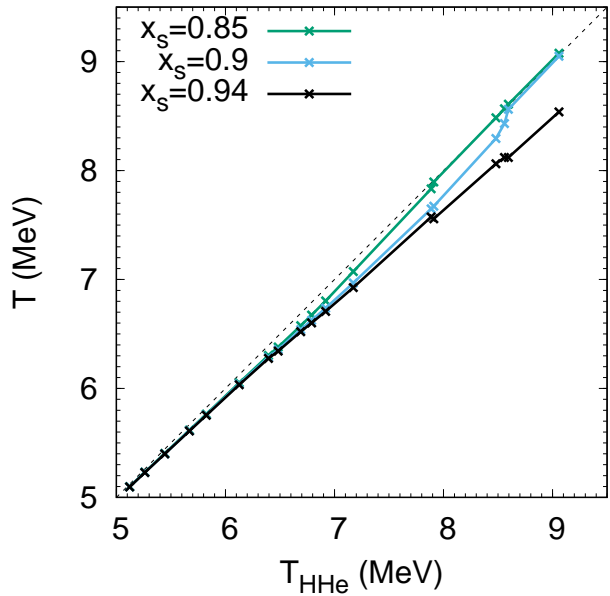}
   \end{tabular}
   \vspace*{-0.8cm}
\caption{$^{124}$Xe$+^{124}$Sn system: Correlation between the input temperature, $T$, of the theoretical model for different $g_{si}$ at $(\rho,y_p)_{\rm exp}$, and the isotopic temperature, $T_{HHe}$ evaluated from the theoretical cluster densities, using experimental values for the density and proton fractions. The dashed line shows $T=T_{\rm HHe}$. }
\label{fig:temp}
\end{figure}

\begin{figure*}
\includegraphics[width=0.9\textwidth]{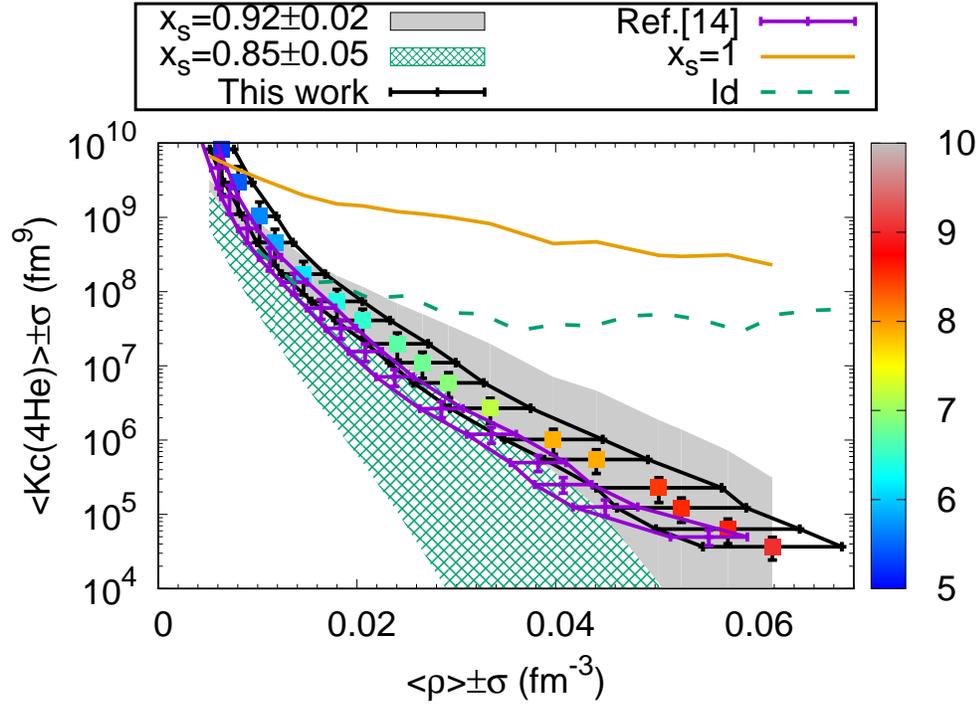}
\vspace*{-0.8cm}
\caption{(Color online) $^{124}$Xe$+^{124}$Sn system: $\alpha$ chemical equilibrium constants as a function of the  density, calculated with the ideal gas prescription for the volume (lower set of data with purple lines)  and uncertainties due to  experimental errors, and with the corrected one (upper set with black lines). In the last case, the uncertainties are due to both the correction and the
  experimental errors. The homogeneous grey and hatched green bands correspond to the chemical equilibrium
  constants from a calculation \cite{PaisPRC2019} where we consider  homogeneous matter with five light clusters, calculated at the  average value of ($T$, $\rho_{\rm exp}$, $Y_{p_{\rm exp}}$), and
  considering different cluster-meson scalar coupling constants $g_{s_i} = x_{s} A_i g_s$. The solid line corresponds to the result with $x_s=1$. The dashed line is the ideal gas result given by Eq. (\ref{eq:ideal}). The color code associated with the data  points represents the temperature in MeV.  }
\label{fig:alphachem}
\end{figure*}  

\begin{figure*}
\includegraphics[width=0.8\textwidth]{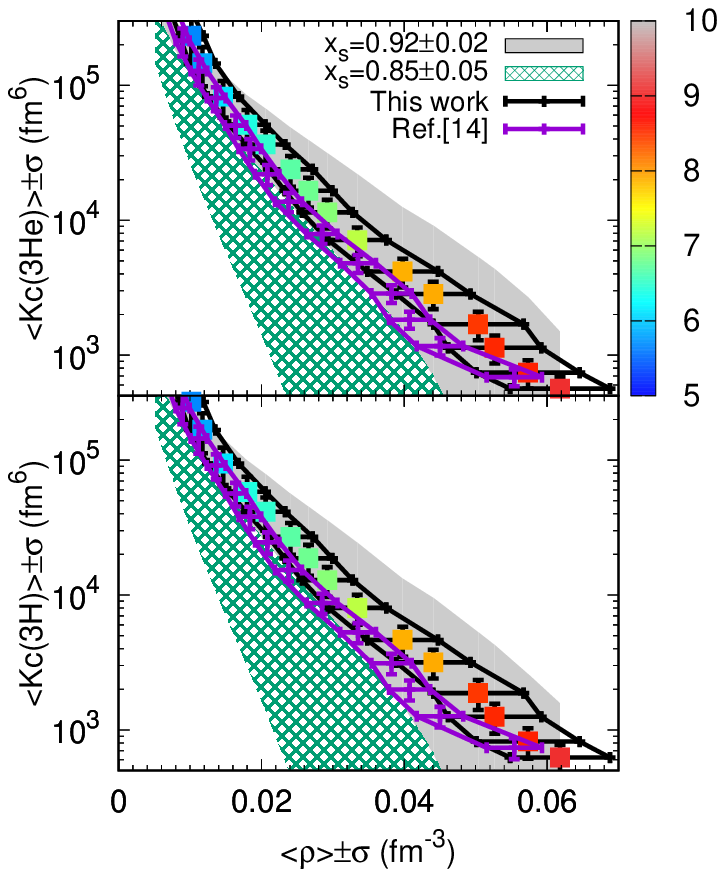}
\caption{(Color online) $^{124}$Xe$+^{124}$Sn system: $^3$He (top) and $^3$H (bottom) chemical equilibrium constants as a function of the  density, calculated with the ideal gas prescription for the volume (lower set of data) and uncertianties due to experimental errors, and with the corrected one (upper set).  In the last case, the uncertainties are due to both the correction and the  experimental errors. The homogeneous grey and hatched green bands correspond to the chemical equilibrium
  constants from a calculation \cite{PaisPRC2019} where we consider
  homogeneous matter with five light clusters, calculated at the
  average value of ($T$, $\rho_{\rm exp}$, $Y_{p_{\rm exp}}$), and
  considering different cluster-meson scalar coupling constants $
  g_{s_i} = x_s A_i g_s$. The color code associated with the data
  points represents the temperature in MeV.   }
\label{fig:3He3Hchem}
\end{figure*}  

\begin{figure*}
\includegraphics[width=0.8\textwidth]{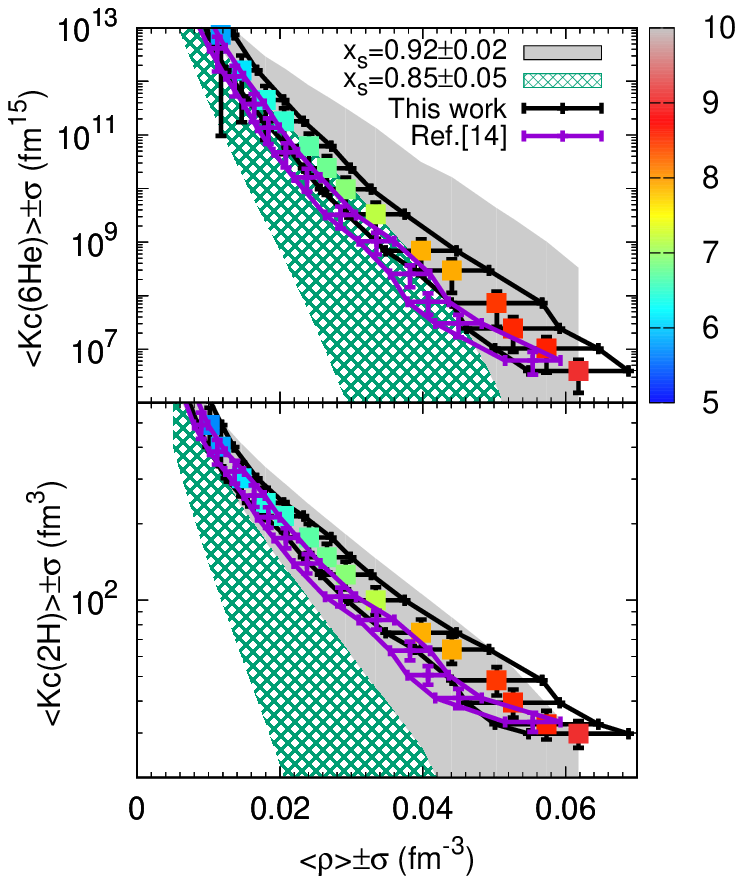}
\caption{(Color online) $^{124}$Xe$+^{124}$Sn system: $^6$He (top) and $^2$H (bottom) chemical equilibrium constants as a function of the  density,  calculated with the ideal gas prescription for the volume (lower set of data) and uncertianties due to experimental errors, and with the corrected one (upper set).  In the last case, the uncertainties are due to both the correction and the  experimental errors. The homogeneous grey and hatched green bands correspond to the chemical equilibrium
  constants from a calculation \cite{PaisPRC2019} where we consider
  homogeneous matter with five light clusters, calculated at the
  average value of ($T$, $\rho_{\rm exp}$, $Y_{p_{\rm exp}}$), and
  considering different cluster-meson scalar coupling constants $
  g_{s_i} = x_s A_i g_s$. The color code associated with the data
  points represents the temperature in MeV. }
\label{fig:6He2Hchem}
\end{figure*}

In Ref.~\cite{PaisPRC97}, a novel approach for the inclusion of
in-medium effects in the equation of state for warm stellar matter
with light clusters was introduced. This model includes a
phenomenological modification in the scalar cluster-meson coupling,
and  includes an extra term in the effective mass of the clusters,
derived in the Thomas-Fermi approximation, which gives rise to
effects similar to the excluded volume approach.
The scalar-meson coupling acting on nucleons bound in cluster $i$ with
mass numer $A_i$ is defined as  $g_{s_i} = x_s A_i g_s$, with $g_s$ the
scalar-meson coupling to nucleons in homogeneous matter and $x_s$ a free parameter. This parameter was fitted in the low-density regime to the Virial EoS, and a value of $x_s=0.85\pm 0.05$ was found.  
  In that work, only four light clusters were considered, $^2$H,
  $^3$H, $^3$He, and $^4$He. Later, in Ref.~\cite{PaisPRC2019},
  heavier light clusters, with $A_{\rm max}=12$, together with an average heavy cluster (pasta phase), were added to the model, as it is expected that heavier unstable clusters also form in neutron rich warm stellar matter. In both works, we compared the chemical equilibrium constants obtained with our model with the NIMROD data \cite{QinPRL108} analyzed assuming an ideal gas expression for the determination of the nuclear density. 
A satisfactory agreement was  obtained for all clusters  but the deuteron using the scalar cluster-meson coupling parameter $x_s = 0.85 \pm 0.05$.

Comparing the model of Refs.~\cite{PaisPRC97,PaisPRC2019} with this new analysis will allow to determine the value of the in-medium parameter $x_s$ in a more consistent way, and at the same time will provide an estimate of the effect of the correction we have introduced, with respect to the analysis method of Refs.\cite{QinPRL108,BougaultJPG19}.  

In order to make this comparison,  the thermodynamic variables
$(T,\rho,y_p)$ corresponding to each $v_{surf}$ bin must be
specified. Among those variables, the density $\rho$ was evaluated
in a model independent way in Section \ref{sec:analysis} under the
constraint that the multiplicities of the different clusters at a
given time of the emission, as measured by the $v_{surf}$ variable,
should correspond to the same density. On the other hand, an ideal gas
assumption was used to determine both the temperature
via Eq.(\ref{eq:temperature}), and the proton fraction
via Eqs.(\ref{eq:yp}),(\ref{eq:rnp}). We have already discussed the fact
that, in the absence of any in-medium correction, the volume extracted
from the $^3$He and $^3$H multiplicities already coincide for all
$v_{surf}$ bins. This is a strong indication that the associated
in-medium effects must be very close, meaning that they necessarily
 cancel in the $y_p$ estimation
Eq.(\ref{eq:rnp}). Moreover, we have also observed that the chemical
constants appear  to be  largely independent of the isospin content of the source (see Fig.\ref{fig5}). 
 An extra confirmation of this hypothesis is given by the fact that we have checked that  eq.(\ref{eq:rnp})
is very well verified in the theoretical model, whatever the value of $x_s$.
For those reasons, we use the experimentally determined $y_p$ value as input of the theoretical model. 

Concerning the temperature, we fix it in each $(\rho,y_p)$ point by imposing that the isotopic thermometer Eq.(\ref{eq:temperature}) evaluated in the model, correctly reproduces the  measured $T_{HHe}$ value.

In Fig.~\ref{fig:temp}, we plot the relation between $T$, the input temperature of the theoretical model,  and the double ratio thermometer response $T_{HHe}$, given by Eq.(\ref{eq:temperature})  using the theoretical particle yields. The density and proton fractions in each point are the ones estimated from the data. Different values for the $x_s$ parameter are considered. 
A deviation  from the $T=T_{\rm HHe}$ limit, given by the dashed line,
is observed  for the largest temperatures, when the density is larger and medium effects more important, and for the highest values of $x_s$, corresponding to larger fractions of clusters.
 This is consistent with the fact that the double ratio thermometer $T_{HHe}$ corresponds to the thermodynamic temperature only in the limit of an ideal gas. However we can see that the correlation between $T$ and $T_{HHe}$ is very good, the in-medium effects largely cancel in the double ratio, and the bias does not exceed 5.75\%. 
 Anticipating our results, the maximum bias observed for the coupling that best reproduces
 the experimental equilibrium constants, $x_s=0.92$, never exceeds 5.75 \%. We have checked that such a bias does not change the experimental results within the error bars.

From this correlation we can extract, for each $x_s$ value, the thermodynamic input sets  $(T,\rho,y_p)$ that are compatible with the $T_{HHe}$.  The resulting chemical constants are compared to the
experimental ones in  Figs.~\ref{fig:alphachem}, \ref{fig:3He3Hchem}, and \ref{fig:6He2Hchem}.

\begin{figure}
  \begin{tabular}{cc}
\includegraphics[width=1\textwidth]{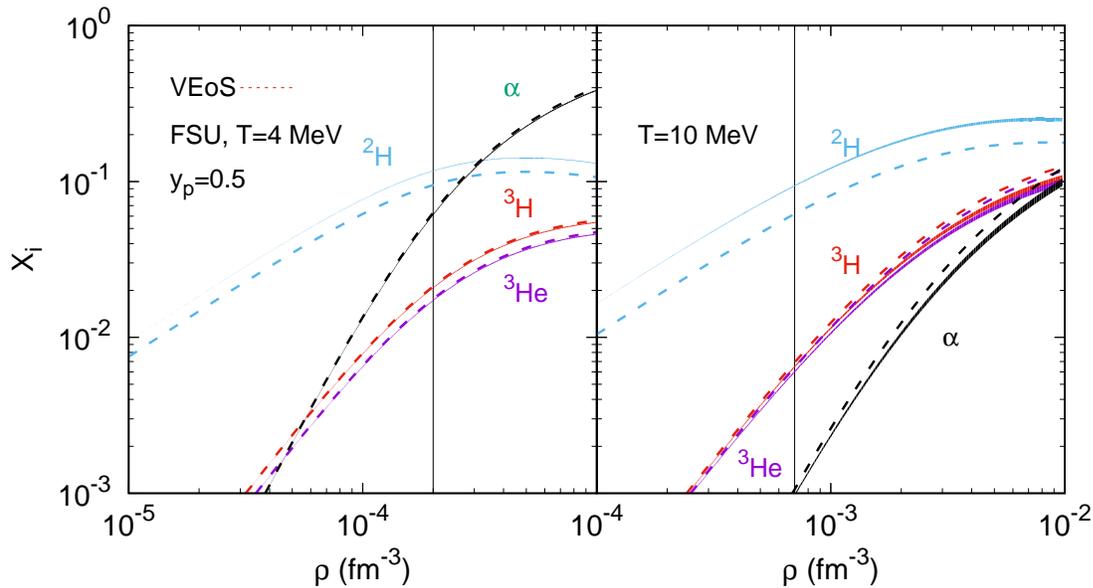} 
  \end{tabular}
\caption{The mass fractions of clusters as a function of the density from a calculation \cite{PaisPRC2019} where we consider homogeneous matter with four light clusters,  and considering different cluster-meson scalar coupling constants $ g_{s_i} = x_s A_i g_s$, with $T=4$ (left) and 10 (right) MeV. The dashed lines are the Virial EoS.}
\label{fig:virial}
\end{figure}  

  The grey (green hatched)  regions are the theoretical
 calculations performed  at different $x_{s}$, and the bands represent the
 experimental quantities.   The  experimental data points are
 represented with different colors that indicate the respective  temperature which changes from 5 to 10 MeV.
 Very similar results are obtained for all the different experimental
entrance channels and therefore were not shown, see Fig. \ref{fig5}.

To appreciate the effect of the correction, the results obtained in the previous analysis \cite{BougaultJPG19}, where the ideal gas hypothesis was used to extract the source volume, are displayed as a lower set of data. We can remark again that the correction induced by this new analysis leads to increased values of both the density and the chemical constants. 
The lower hatched bands in the figures correspond to the calculation of Ref.\cite{PaisPRC97}, where the in-medium corrections were optimized to the uncorrected data of Ref.\cite{QinPRL108}.

 We can see that we need higher values of $x_s$ in order to fit the data, with respect to the results of Refs.~\cite{PaisPRC97}, ~\cite{PaisPRC2019}, and  an optimal value can
be extracted as $x_s=0.92\pm 0.02$, which is seen to produce an
adequate reproduction of the whole set of experimental data. This means
 that the nucleons in the clusters behave closer to the unbound nucleons, which correspond to $x_s=1$, than previously deduced.

Still, we can observe a slight mass effect which does not seem fully accounted by the present calculation. Indeed, the optimal $x_s$ value tends to overestimate the equilibrium constant of the heaviest species ($^6$He) and underestimate the one of the lightest cluster ($^2$H). This might suggest that the hypothesis of the model, namely the fact that the coupling to the meson fields scales linearly with the number of nucleons bound in each cluster, could be not fully correct, and a more ab-initio treatment would be in order. This point is left for future work.

It is also interesting to observe that the uncorrected data set of Ref.\cite{BougaultJPG19} (lower data set in Figs.~\ref{fig:alphachem} - \ref{fig:6He2Hchem}) appear only marginally compatible with the previous estimation $x_s=0.85\pm 0.05$.
Indeed, the estimation $x_s=0.85\pm 0.05$, which nicely reproduced the uncorrected Qin et al. data, cannot describe neither the deuteron, nor the $A=3$ isobars of this new data set. Conversely, the introduction of in-medium effects allows a simultaneous reproduction of the whole data. To confirm the new value of the universal coupling  $x_s=0.92\pm 0.02$, and at the same time test the compatibility of the different data sets, it would be very interesting to apply this new analysis to the NIMROD data, such as to verify if the same value for the $x_s$ parameter is able to reproduce both sets of data, once the source volume is correctly estimated from the Bayesian analysis.

At very low density, some constraints on the in-medium binding energy shifts can be obtained from the ab-initio virial equation of state. The results of our model, with the value of $x_s$ optimized on the new analysis of the chemical constants, is compared to the virial constraint in Fig.~ \ref{fig:virial}. We can see that the new estimation of the $x_s$ parameter is still within the virial EoS limits. A larger scalar-meson coupling  has
  strong implications on the dissolution density of the clusters: 
  clusters will survive at larger densities.

\section{Conclusions} \label{sec:conc}

A new analysis, where in-medium effects were included via a correction
to the internal partition function, was done to the experimental data
on the formation of nuclear clusters in heavy-ion collisions
presented in Ref.~\cite{BougaultJPG19}. This was done by including a
 global effective correction to the binding energy of the cluster.

Using a Bayesian analysis, the probability distribution for the parameters of the in-medium
correction was determined by imposing that the free volumes obtained for the
different clusters in a given source  velocity bin be
compatible. The main consequence of including the in-medium
correction  was to reduce the free volume, and, as a
consequence, increase  the density.

In the  absence of any in-medium correction,
only the volume extracted from the $^3$He and $^3$H multiplicities 
coincide for all source velocity bins. This indicates that the associated
in-medium effects are similar and cancel in the calculation of the
proton fraction, and suggests that, for a given thermodynamic condition and a given cluster mass, the correction should not 
strongly depend on the charge of the clusters. Besides, the equilibrium constants obtained showed to
be quite independent of the isospin content of the three systems that
were analysed. It would be interesting to check this point with a larger isospin
range.

Different functional forms for the correction as a function of the mass and isospin of the cluster were considered, and we observed that the results, both
 for the thermodynamic conditions and chemical equilibrium  constants,
 were similar. 

 A comparison to a theoretical model \cite{PaisPRC97,PaisPRC2019} was
  also done. The modification of the experimental chemical constants 
due to the better evaluation of the source density leads to a modified estimation for
the magnitude of the scalar-meson
  coupling $x_s$ for nucleons bound in clusters:  the optimal values extracted are $x_s = 0.92 \pm
  0.02$, larger than the ones  found in  \cite{PaisPRC97},
  $x_s=0.85\pm 0.05$. Values of $x_s$ closer to the unity mean that
  the effect of the medium is less important than previously estimated, and that clusters will melt at larger densities. In turn, this means that the contribution of clusters in the neutrino opacity in the deleptonization phase of the proto-neutron star should  be more important than previously considered \cite{Fischer}.

In a future work, it would be extremely interesting to calculate explicitly the neutrino opacity in the relevant thermodynamic conditions. Prior to that, it will be important to perform a new analysis of the experimental data of Ref.~\cite{QinPRL108}, including the possibility of in-medium effects in the determination of the effective volume  in the same spirit as the one presented in this paper, in order to check the consistency of the different data sets, and to settle the model dependence of the results.

\ack{
This work was partly supported by the FCT (Portugal) Projects No. UID/FIS/04564/2019, UID/FIS/04564/2020 and POCI-01-0145-FEDER-029912, and by PHAROS COST Action CA16214. H.P. acknowledges the grant CEECIND/03092/2017 (FCT, Portugal). She is very thankful to F.G. and her group at LPC (Caen) for the kind hospitality during her stay there within a PHAROS STSM, where this work started. This work is part of the INDRA collaboration program. We thank the GANIL staff for providing us the beams and for the technical support during the experiment. We acknowledge support from R\'egion Normandie under RIN/FIDNEOS.}

\section*{References}
\bibliographystyle{iopart-num}

\end{document}